# PERSPECTIVE

# Bridging the gap between photovoltaics R&D and manufacturing with data-driven optimization

**Authors:** Felipe Oviedo[1*&], Zekun Ren[2,3*], Xue Hansong[3], Siyu Isaac Parker Tian[2,3], Kaicheng Zhang[4], Mariya Layurova[1], Thomas Heumueller[4,5], Ning Li[4,5,6], Erik Birgersson[7], Shijing Sun[1], Benji Mayurama[8], Ian Marius Peters[1,5], Christoph J. Brabec[4,5], John Fisher III[1], Tonio Buonassisi[1,2]

*These authors contributed equally.
&Corresponding authors.
[1]Massachusetts Institute of Technology, 77 Massachusetts Ave., Cambridge MA 02139, [2]Singapore-MIT Alliance for Research and Technology SMART, Singapore 138602, Singapore, [3]Solar Energy Research Institute of Singapore (SERIS), National University of Singapore, Singapore 117574, Singapore, [4]Institute of Materials for Electronics and Energy Technology (i-MEET), Friedrich-Alexander University Erlangen-Nürnberg, 91058 Erlangen, Germany, [5]Helmholtz Institute HI-ErN, Forschungszentrum Jülich, Immerwahrstr. 2, 91058 Erlangen, Germany, [6]National Engineering Research Center for Advanced Polymer Processing Technology, Zhengzhou University, 450002 Zhengzhou, China, [7]National University of Singapore, Singapore 117574, Singapore, [8]Airforce Research Laboratory (AFRL), Dayton, Ohio.

**Context and Scale**

Novel photovoltaics, such as perovskites and perovskite-inspired materials, have shown great promise due to high efficiency and potentially low manufacturing cost. So far, solar cell R&D has mostly focused on achieving record efficiencies, a process that often results in small batches, large variance, and limited understanding of the physical causes of underperformance. This approach is intensive in time and resources, and ignores many relevant factors for industrial production, particularly the need for high reproducibility and high manufacturing yield, and the accompanying need of physical insights. The record-efficiency paradigm is effective in early-stage R&D, but becomes unsuitable for industrial translation, requiring a repetition of the optimization procedure in the industrial setting. This mismatch between optimization objectives, combined with the complexity of physical root-cause analysis, contributes to decade-long timelines to transfer new technologies into the market. Based on recent machine learning and technoeconomic advances, our perspective articulates a data-driven optimization framework to bridge R&D and manufacturing optimization approaches. We extend the maximum-efficiency optimization paradigm by considering two additional dimensions: a technoeconomic figure of merit and scalable physical inference. Our framework naturally aligns different stages of technology development with shared optimization objectives, and accelerates the optimization process by providing physical insights.

**Summary**

Solar cell R&D and industrial production have mismatched objectives and approaches, which result in poor and lengthy industry transfer of novel photovoltaic technologies. In this perspective, we propose a new data-driven optimization framework to bridge this divide. Our framework incorporates both technoeconomic analysis and scalable physical inference, a machine-learning approach to elucidate root cause(s) of underperformance. First, we propose a manufacturing-relevant figure of merit, total revenue, which combines efficiency maximization with manufacturing variability and yield. Our figure of merit allows a smooth transition from R&D to industrial production. Subsequently, we propose scalable physical inference, informed by a device model, to estimate the impact of solar cell parameters on a given figure of merit. Our parameter inference method accelerates root-causes analysis and provides key physical insights for the optimization process. As a case study, we fabricate 144 perovskite devices, and contrast our framework to traditional maximum-efficiency optimization.

**Introduction:**

Traditionally, early-stage photovoltaics research and development (R&D) has focused on optimization for record efficiencies[1]. When applied to highly-reproducible and well-understood material systems, such as silicon or III-V materials, this efficiency-guided optimization paradigm has allowed successful translation to industrial production. In contrast, novel photovoltaics, such as organics, perovskites or perovskite-inspired materials, often lack reproducibility due to very sensitive chemical synthesis and crystallization mechanisms [2–8], challenging process control [9–12], and complex interactions among compositional elements [2,13–15]. In this high-variability context, maximizing solar cell efficiency is often overemphasized at the expense of factors that impact industrial production and economic revenue, such as process reproducibility and manufacturing yield. This record-efficiency focus has scientific and economic justification in the early stages of research; however, as a technology matures, failure to abandon this single-objective paradigm can delay industrial adoption.

At the device level, hypothesis testing and underperformance root-cause analysis has, traditionally, been accomplished with dedicated samples and in-depth characterization techniques. In the case of novel photovoltaics, as manufacturing-relevant factors become more important, this traditional approach does not scale well due to large variability when sample numbers increase, reduced process control and difficulty to relate figures of merit to physical parameters. In consequence, two main issues affect the transition to industrial production: (1) the mismatch of R&D and industrial manufacturing figures of merit, and (2) the increasing complexity of hypothesis testing and root-cause analysis with larger batch sizes. These factors contribute to decade-long timelines for technology transfer, as promising photovoltaic technologies need to be re-optimized for industrial production with significant effort and resources.

Commonly, solar cells have been optimized through various methods including: restricted parameter sweeps [8], expert-guided trial-and-error, design of experiments [16] and process control [9,17,18]. This traditional optimization paradigm relies on significant experimentalist expertise, along with time-consuming and complex characterization methods to extract physical root-causes of underperformance or perform hypothesis testing. As high-throughput material development becomes more popular [19–22], methods that once were reserved for industrial settings and mature technologies have become available earlier in the R&D cycle. These methods, such as automated solution synthesis or high-throughput characterization, allow rapid experimental screening of vast parameter spaces, and produce a distribution of multiple sample measurements at each variable condition. Thanks to these improvements, recent approaches have demonstrated the use of machine learning methods to model the variable space and identify optimal variable combinations [16,20,23], or extract root-causes of underperformance by using an auxiliary physical model[24]. In addition, modern technoeconomic analysis can inform solar cell optimization priorities[25–27].

In this perspective, we propose a data-driven framework to bridge the divide between R&D and industrial manufacturing, by integrating several emergent quantitative tools: (1) **Technoeconomic analysis** can define technoeconomic optimization targets, which balance efficiency and manufacturing factors at various stages of the development cycle; (2) **Scalable physical inference** based on a device model can extract bulk and surface solar cell parameters (*"device parameters"*) that directly affect performance, accelerating hypothesis testing and root-cause analysis. We integrate these tools into a data-driven optimization framework. Figure 1 presents a conceptual comparison of the traditional maximum efficiency paradigm, and our proposed approach to integrate technoeconomic and physical inference considerations. We detail this framework in the next sections, perform a case study of its application to perovskite devices, and outline the opportunities and challenges for generalizing our proposed approach.

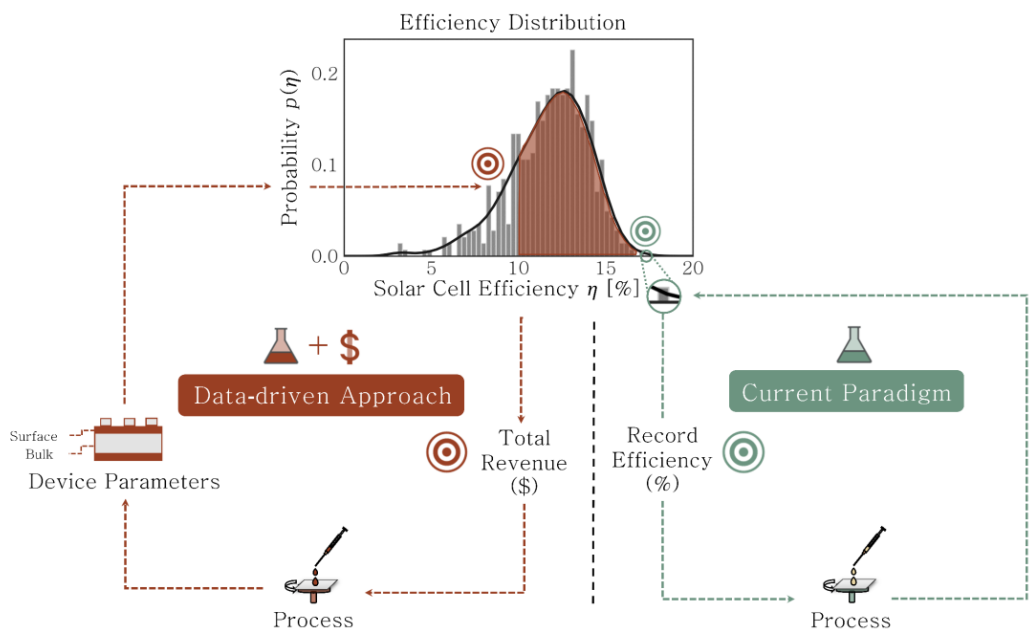

**Figure 1**: **Conceptual comparison of the traditional R&D paradigm (green) and our data-driven optimization approach (red).** The current R&D paradigm tries to maximize solar cell efficiency with the aim of reaching record efficiencies, with limited interest in optimizing the efficiency distribution or the device parameters, *i.e.* bulk and interface properties that govern device performance. In our data-driven framework, we extend the current paradigm by considering technoeconomic factors, linked to the shape of the efficiency distribution, and physical device parameter inference. This framework has the potential to accelerate the translation between R&D and industrial manufacturing.

**Data-driven Optimization Framework**

Figure 2A details the flow of: I. the traditional R&D optimization process, and II. our proposed framework for revenue optimization. In the R&D optimization paradigm, samples are fabricated at given conditions $\theta$, which could be process-related or device-related, and the optimal conditions $\theta^*$, producing the absolute maximum efficiency $\eta^*$, are chosen. In practice, the best result is reported and, if it is a record solar cell efficiency, is certified. Hypothesis testing and root-cause analysis are performed using in-depth characterization techniques, helping to elucidate factors conducive to high-performing solar cells. In some cases, especially in reproducibility-focused studies, the mean efficiency is optimized. Paradigm I. is functional in the early stages of the R&D cycle: when the efficiency of a single best-performing device has significantly more value than the rest of efficiencies in the fabrication batch because it best represents the potential performance of the corresponding experimental condition. In this case, maximizing single cell efficiency, regardless of batch variance, could lead to a faster learning rate than optimizing every cell from the batch. Naturally, as a technology matures, and we move from scientific discovery to industrial application, the efficiency distribution becomes more important, as the standard deviation and manufacturing yield become targets for optimization. Optimizing for maximum or mean efficiency does not consistently integrate these manufacturing-relevant considerations. Furthermore, while root-cause analysis and hypothesis testing are still essential activities during optimization, the larger batch size makes characterization more time-consuming or infeasible.

To address these issues, we propose our data-driven optimization framework, II. in Figure 2A. This framework relates optimization conditions to total revenue $R_T$, using emerging quantitative tools. A batch of samples produces a probability distribution of efficiencies at given conditions $\theta$, defined as $p(\eta|\theta)$. We propose to use a technoeconomic efficiency-revenue function $f_R$, shown in Figure 2B, applied over each sample in the efficiency probability distribution, to compute the total revenue statistic $R_T$. This new optimization objective aims to better balance the relative value of the efficiency distribution. Once computed, a machine learning model can be used to map process conditions to $R_T$, in order to approximate $R_T$ in unexplored regions of the variable space. We can use surrogate-based optimization, such as Bayesian optimization[28], to find the optimal value of $\theta^*$ which maximizes the expectation of $R_T$. Finally, scalable device parameter inference, relying on high-throughput measurements and a calibrated physical model, can reduce the time and complexity of root-cause analysis, and allow targeted parameter optimization.

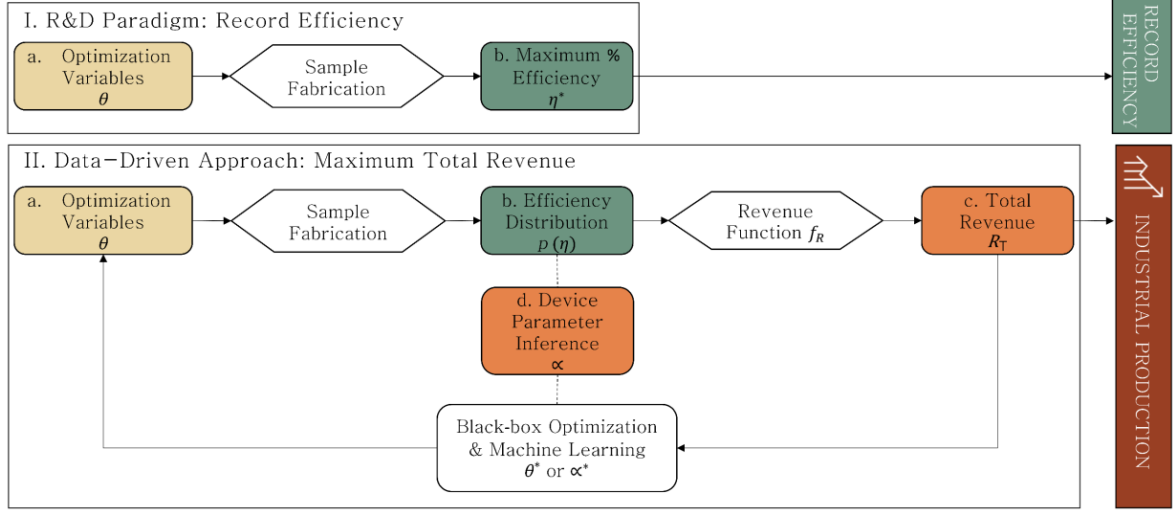

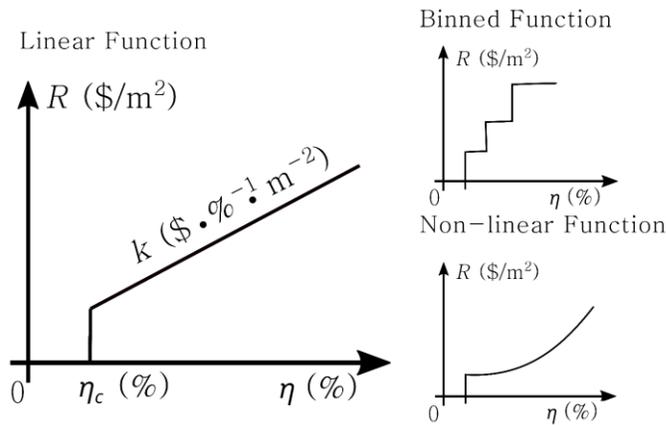

*Figure 2: A) **Traditional R&D Maximum-Efficiency Optimization vs Proposed Total Revenue Optimization.** Traditional Optimization (I) leads to record device efficiencies, while (II) Total Revenue Optimization leads to industrial performing devices with adequate trade-off of efficiency, variability and yield. B) **Various embodiments of the technoeconomic efficiency-revenue function $f_R$,** where $k$ corresponds to the marginal value of efficiency and $\eta_C$ is the minimum solar cell efficiency (%) that is market-competitive.*

**Efficiency-Revenue Function $f_R$ and Technoeconomic Optimization Objective**

A manufacturing-relevant optimization objective must consider the distribution of the efficiencies of the samples in a batch, rather than a single value. This requirement allows embedding of variability, which we define as the standard deviation of the efficiency distribution, along with manufacturing yield in the optimization. For this purpose, a multitude of optimization objectives can be constructed, *e.g.* the ratio of mean value and standard deviation of efficiency in a fabrication batch. These figures of merit are often arbitrary and depend on the specific efficiency distribution. In contrast, rather than focusing on purely technological factors, we propose to define the optimization objective based in technoeconomic considerations. We define *total revenue* $R_T$ as the cumulative revenue, in $, of producing a batch of solar cells at a given condition. To map the distribution of sample efficiencies $\eta$ to revenue $R$, we define a technoeconomic *efficiency-revenue function* $f_R$ as a truncated linear function that approximates the marginal value of efficiency per % efficiency point, such that:

$$f_R(\eta) = k * \eta \text{ if } \eta \geq \eta_c,$$
$$f_R(\eta) = 0 \text{ if } \eta < \eta_c \qquad [1]$$

where $k$, in units of $/(m^2 \cdot \%)$, is the slope of the linear efficiency-revenue function, and $\eta_C$ is the minimum solar cell efficiency, in %, that is market-competitive, or below which the economic value of the solar cells is negligible. A schematic of this linear $f_R$ is shown in Figure 1B.

The choice of $f_R$ can be made according to available information, along with the stage and priorities of the photovoltaics development process. In one reasonable approach, within a few percent absolute efficiency variation around the market-average efficiency, $f_R$ can be approximated as a linear function, with slope $k$ equivalent to the prevailing market rate or, alternatively, the minimum sustainable price. In the latter case, the units of $/W can be transformed to $k$ units by assuming 1-sun conditions and a given solar cell area. The value of $k$ is easily discoverable, given market data for mature crystalline silicon and thin-film selling prices [29–32], or projected costs of novel photovoltaics. This linear, first-order approximation to $f_R$ is not always correct: as one deviates significantly from market-average efficiencies, non-linearity in $f_R$ is expected[31], and $f_R$ discovery becomes challenging due to limited publicly-available market data. Similarly, $f_R$ can be comprised of several small step functions, reflecting the "efficiency bins" in an actual production line. Figure 1B illustrates these alternative functions. An important assumption for $f_R$ is that the cost structure and capital expenditure of the fabrication process are comparable between R&D and various scales of manufacturing. Thus, for early-stage and some maturing technologies we could expect $f_R$ to change as process costs change. According to the development stage, $f_R$ can be sequentially adjusted or reweighted to account for the shortcomings listed above. For instance, we can combine two functional shapes for $f_R$ as efficiency, or manufacturing processes, change. Finally, we further modify $f_R$, setting the economic zero at efficiencies below the critical cut-off efficiency $\eta_c$. This reflects the fact that efficiencies below a critical threshold cannot be binned and sold during industrial production. This quantity can also be calculated from levelized cost of energy calculations, module degradation rates and system lifetime[27].

Once $f_R$ is defined, we can compute the mean revenue $\bar{R}$ at a given process condition by taking the mean of $f_R$ over the efficiency probability distribution, for a given solar cell area:

$$\bar{R} = \int f_R(\eta) \, p(\eta|\theta) \, d\eta \quad [2]$$

Eq. 2 utilizes the probability density function (pdf) of $p\,(\eta|\theta)$, which is not known perfectly. However, there are a variety of ways to estimate $\bar{R}$ ranging from a simple empirical average, to use of parametric densities under an estimate of the parameters, to a kernel density estimate (and more) [33]. Errors in the estimation of $p\,(\eta|\theta)$ could easily affect the final optimization objective. A reasonable approach makes use of the empirical distribution of the fabrication data to estimate the mean revenue as a statistic. Based on $\bar{R}$, we compute the total revenue at a given condition $R_T$. If we assume cell area is the same for all cells, we can use the number of samples $N_\theta$ at condition $\theta$ such that:

$$R_\mathrm{T} = N_\theta * \bar{R} = \sum_{i=1}^{N_\theta} f_\mathrm{R}(\eta_i) \quad [3]$$

Although equivalent to $\bar{R}$, $R_\mathrm{T}$ constitutes an intuitive and manufacturing-relevant figure of merit, as it is the final figure for estimating the top-line of a given technology for a certain batch size, and is consistent with various modifications of $f_\mathrm{R}$, such as step-wise binning.

To illustrate how $R_\mathrm{T}$ balances efficiency, variability and manufacturing yield, two simulated efficiency distributions with the same non-zero mean efficiency are presented in Figure 3. The distributions represent opposite stages in the R&D and industrial production spectrum. Figure 3A, illustrates an R&D distribution: a zero-inflated normal distribution, simulating early-stage research with high-variability, many failed samples (zero-inflation) and some high-efficiency samples (over 19%). Figure 3B illustrates a manufacturing distribution: an inverted log-normal distribution with small standard deviation[34,35], corresponding to a manufacturing process which has high reproducibility but low probability of surpassing certain efficiency. Analytical expression of these distributions are included in the SI. In figures 3C and 3D, we compute $R_\mathrm{T}$ for each distribution, normalized by the total simulated number of cells, using varying values of $k$ and the ratio between $\eta_\mathrm{c}$ and the non-zero mean efficiency $\eta_\mathrm{mean}$. Several regimes become identifiable. In Figure 3A and the corresponding Figure 3C, the change in $\eta_\mathrm{c}/\eta_\mathrm{mean}$ has a smaller impact on $R_\mathrm{T}$ than in the case of Figure 3D. Thus, variability in R&D distributions has a positive impact if the distribution reaches large enough efficiencies, and the slope $k$ of the efficiency-revenue function is high. This is consistent with the goals of early-stage R&D: optimizing for maximum efficiency, which is in general much more valuable than any other efficiency in the distribution. However, as we move to manufacturing distributions that have lower variability, such as Figure 3B, we observe that a higher $R_\mathrm{T}$ value is achieved by having $\eta_\mathrm{mean}$ above $\eta_\mathrm{c}$. In addition, $k$ has a larger impact on $R_\mathrm{T}$ in the manufacturing distribution than in the R&D distribution, as can be seen by comparing Figure 3D and Figure 3C. These trends are consistent with the manufacturing goals of minimizing variability and increasing manufacturing yield. From this analysis, we can conclude that, given a choice of $k$ and $\eta_\mathrm{c}$ relative to the distribution, the target objective $R_\mathrm{T}$ is consistent with both motivations of R&D and industrial production, addressing the objective mismatch between both. In consequence, using $R_\mathrm{T}$ early in the transition from R&D and to industrial production can be beneficial.

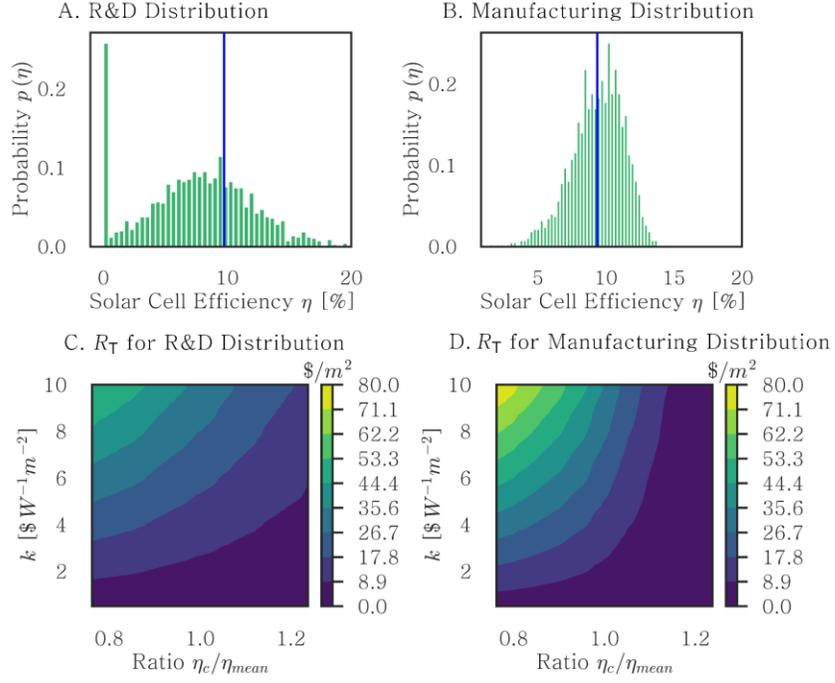

*Figure 3:* A. Simulated R&D Efficiency Distribution, B. Simulated Manufacturing Efficiency Distribution, C. Total Revenue Contour Plot for R&D Distribution, D. Total Revenue Contour Plot for Manufacturing Distribution.

**Scalable Physical Inference and Optimization Loop**

In this perspective, in addition to the technoeconomic objective $R_T$, we propose to use scalable device parameter inference to accelerate solar cell optimization. The combination of these two fundamental contributions into a machine learning framework allows us to: 1) extrapolate solar cell figures of merit to unexplored regions of the parameter space, 2) provide physical insights of underperformance and confirm or discard solar cell hypothesis, and 3) allow targeted optimization of device parameters.

To formalize this approach, we formulate solar cell improvement as the optimization of a black-box function $f(\theta)$, such that:

$$\theta^* = \underset{\theta}{\mathrm{argmax}}\, \mathbb{E}[f(\theta)] \quad [4]$$
$$f(\theta) = g(\alpha(\theta)) \quad [5]$$

The variable $\theta$ corresponds to device and process conditions, such as film annealing temperature or active layer composition, which we control directly. $\theta^*$ is the optimal set of variables that maximize the mean of the objective. $f(\theta)$ is a noisy black-box function of solar cell efficiency, or other performance metric, which we evaluate by fabricating and measuring samples. Nevertheless, the relation between $\theta$ and the final performance $f(\theta)$ is often difficult to interpret physically, causing difficulties in decision-making for the experimentalist. We address this issue by reformulating $f(\theta)$ as $g(\alpha(\theta))$. The device parameters $\alpha(\theta)$, such as bulk lifetime or surface recombination velocity, determine the final performance of a solar cell, and have physical significance according to a physical model of solar cell operation given by $g(\alpha(\theta))$.

Traditionally, the latent parameters $\alpha(\theta)$ are estimated by dedicated auxiliary samples and time-consuming secondary characterization techniques, *e.g.* time-resolved photoluminescence measurements, or by performing statistical inference (fitting) of a forward device numerical model of $g(\alpha(\theta))$[36,37]. The latter approach is often computationally expensive and is not scalable to a large number of samples. Recently, rapid electric solar cell characterization and machine learning techniques have made possible, if a reliable physical model $g(\alpha(\theta))$ is available, to perform scalable statistical inference in a large number of solar cell samples[38,39]. Surrogate machine learning models, such as neural networks trained on numerical device models, have been applied to compute physical models orders of magnitude faster. These techniques have proven useful to quickly identify root-cause(s) of underperformance[24], and accelerate the optimization process[23].

As shown in Figure 1A, the optimization allows us to find optimal device parameters $\alpha^*$, such that:

$$\alpha^* = \operatorname*{argmax}_{\alpha} \mathbb{E}[g(\alpha)] \quad [6]$$

Or, alternatively, to determine the process conditions $\theta$ that selectively optimize (either maximize or minimize) a particular device parameter $\alpha_i$:

$$\theta^* = \operatorname*{argmax}_{\theta} \ \mathbb{E}[\alpha_i(\theta)] \text{ or } \operatorname*{argmin}_{\theta} \ \mathbb{E}[\alpha_i(\theta)] \quad [7]$$

The 'inner' optimization loops in Eq. 6 and Eq. 7 can be used to perform targeted optimization of device parameters, or gain physical insights to inform optimization decisions. Depending on the figure of merit, this allows to optimize device parameters for particular objectives, and balance tradeoffs that are often found in manufacturing. For example, we could use the 'inner' optimization loop to determine high-priority physical parameters for maximizing manufacturing yield (Eq. 6), or choose process conditions that improve surface passivation by minimizing surface recombination velocity (Eq. 7).

Once the optimization problem is defined, there are multiple ways to find the optimum, including expert heuristics, auxiliary characterization, parameter sweeps, design of experiments or surrogate-based optimization. Among the most sample-efficient methods[28], surrogate-based optimization aims to map the response surface of the variable space to identify the optimum. This approach includes methods that are both open-loop and closed-loop, such as response surface method, Bayesian optimization, etc. In this context, the proper mapping of the parameter space using the available samples becomes a critical part of the optimization process. Multiple supervised machine learning methods have been employed in the past, including support vector regression [16], Gaussian process regression [40], Bayesian neural networks [41] and Bayesian networks[23].

**Case Study: Perovskite Optimization**

To illustrate our approach, we choose Methylammonium Lead Iodide (MAPbI₃), as it is a representative example of a novel photovoltaic technology currently transitioning to industrial manufacturing. We fabricate 144 perovskite solar cells at 12 different combinations of two dominant process variables: solvent ratio of the DMF:NMP (Dimethylformamide:N-Methyl-2-pyrrolidone) precursor solution and film annealing temperature. The device architecture, along with fabrication details, are included in the SI. We vary the annealing temperature in the range of 70 – 130 °C and the solvent ratio of DMF to NMP in the range of 2 – 8. The chosen process variables and their magnitude are determinant of the film quality during fabrication, and affect both bulk and interfacial properties of the devices after fabrication[7,42,43]. Annealing temperature determines the grain and film morphology, with higher annealing temperature being favorable for solvent removal and large grain formation, but making the samples more prone to decomposition and in-film variability[44]. Solvent engineering, on the other hand, determines the film uniformity by forming intermediate phases of lead-solvent-complexes[45]. Both the annealing temperature and solvent ratios have direct influence on the kinetic and thermodynamic parameters that govern the crystallization of perovskite phases on the substrates[43,45]. Traditionally, both variables are chosen to maximize either the highest efficiency or, less commonly, the mean efficiency of the resulting solar cells. Often, the optimization is time-consuming, and requires sweeping each variable independently or performing a grid search of the variable space. For each fabricated cell, we measure the current-voltage characteristics at 10 illumination conditions, between dark and 5 suns, and compute the AM1.5G 1 Sun solar cell efficiency. Table S1 summarizes the AM1.5G maximum efficiency, mean efficiency, variability and yield (with a $\eta_c = 12\%$) for each process condition. The corresponding solar cell efficiency distributions are presented in Figure S2. The observed broad variability at a single condition has various causes, including non-linear sensitivity to crystallization conditions, in-substrate variation caused by non-controllable factors, such as seasonal humidity or batch variation of raw materials, as well as human error[46].

We make several assumptions about the $f_R$ of perovskites, according to cost modeling in [29,30]. We choose $k = 0.3\ \$/W = 3\ \$/(m^2 \cdot \%)$, which is close to the estimated minimum sustainable price for perovskites solar cells in [30] and [29]. We assume that the chosen $k$ is similar to $k$ of our fabrication process. In the same way, we set $\eta_c$ as 12% efficiency. This value is an approximation based on research-quality solar cells having lower mean efficiency than industrial cells. We include a sensitivity analysis of our assumptions in section III of the SI. After optimization objectives are computed, we use Gaussian process regression (GPR) to interpolate across the variable space. Section IV of the SI describes the model selection procedure. The final chosen model is a homoscedastic GPR with the Matérn 32 kernel[40], which allows a good compromise between smoothness and model capacity.

Figure 4 presents the final mean of the GPR-interpolated space for each optimization objective, along with the location of the optimal process conditions found in open-loop fashion. The white-dotted contours in Figure 4 correspond to 5% contour lines away from the optimum. For each one of the 12 process conditions of interest, shown in the figure as light semi-transparent circles, we compute: maximum efficiency, variability and yield with 12% $\eta_c$. Figures 4A, 4B and 4C show respectively each one of these variable spaces.  In the same way, we compute the total revenue objective, using $k = 0.3\ \$/W =$ 3 $/($m^2 \cdot$ %), with cutoff $\eta_c$ of 0% in Figure 3D, and 12% in Figure 4E. Although the process conditions are relatively coarse, we observe that maximum efficiency and variability have optima at very different locations in the parameter space. Figure 4D successfully combines both objectives, as the optimum's location is intermediate to both maximum efficiency and minimum variability optima. In similar way, when choosing $\eta_c$ to be above 0% in Figure 4E, we find that yield diminishes significantly the values of $R_T$, and affect the optimum's location in comparison to Figure 4D. From these results, we conclude that $R_T$ satisfactorily combines the magnitude and variability of the efficiency distribution, along with yield considerations.

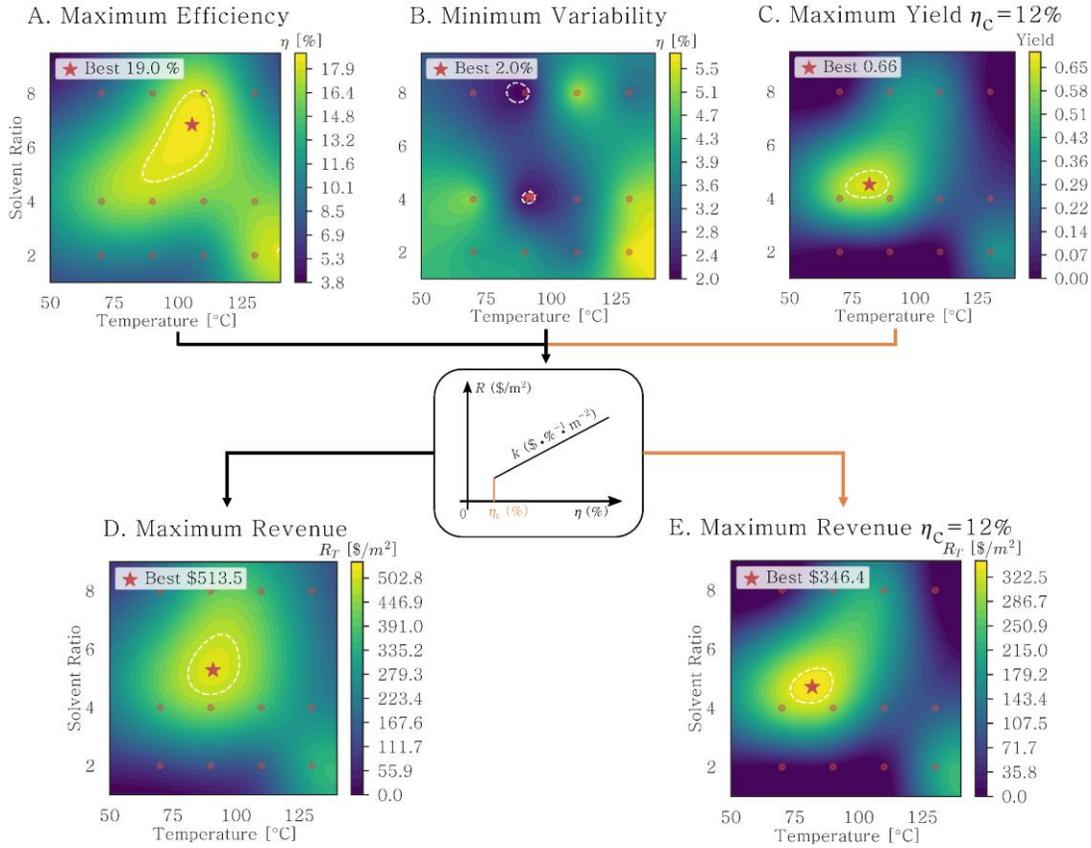

*Figure 4: Optimal process conditions and variable space for various optimization objectives. A. Maximum Efficiency, B. Minimum Variability (Std. Deviation), C. Total Revenue with $\eta_C = 0\%$, which combines the effects of maximum efficiency and variability, D. Total Revenue with $\eta_C = 0\%$, which combines the effects of efficiency, yield and variability, E. Total Revenue with $\eta_C = 12\%$. White-dotted lines are 10% contours away from the optimum, and orange dots represent experimental process conditions.*

According to our quantitative framework, we extract the device parameters $\alpha(\theta)$ of each solar cell to gain physical insights and solve the optimization procedure in Eq. 7. For this purpose, we use a well-calibrated perovskite device model[37]. The forward model solves drift-diffusion partial differential equations, and outputs JV curves at varying illuminations[24] for given device parameters. However, the numerical fitting of such a large number of JV curves is challenging. Here, we use a convolutional neural network (CNN) as a surrogate model to invert the forward device model. To train the CNN, we first generate 20,000 synthetic

JV curves under different illumination intensites using the device model of [37], using random device parameter values sampled from an uniform distribution. The device parameters used to evalue the model are bulk lifetime ($\tau$), front and rear surface recombination velocity ($FSRV$, $RSRV$), series resistance ($R_s$) and shunt resistance ($R_{sh}$). These variables allow to calculate JV curves at different illumination intensitites. Finally, the experimental JV characteristics are fed into the trained CNN to extract the material parameters for each one of the 144 cells. The CNN architecture and its performance, along with details of the perovskite forward model, are included in the SI.

Due to the wide variance of efficiency distributions and the intrinsic model limitations, the inferred device parameters can vary widely, even for a single process condition, as presented in tables S2 and S3. To get useful insights, we compute the *mean* device parameters at the optimal process conditions. In order to do this, we use kernel ridge regression[33], which can be seen as a frequentist version of GPR, to model the mean of each device parameter distribution $\mathbb{E}[\alpha_i(\theta)]$ as function of process conditions:

$$\mathbb{E}[\alpha_i(\theta)] = f(\theta) \quad [8]$$

Hereafter, we interpolate the mean material parameters for each optimal process condition according to the specific optimization objective. Figure 5A illustrates this process. Figure 5B presents the inferred, normalized mean device parameters at the optimal process conditions, considering two cases: optimization for maximum efficiency, $\eta_{max}$, and optimization for maximum total revenue $R_T^*$. It is evident that, on average, $\bar{R}_{sh}$ and $\bar{\tau}$ are higher for solar cells at the $R_T^*$ optimum, compared to cells at the $\eta^*$. In this situation, maximizing revenue entails maximizing $\bar{R}_{sh}$ and $\bar{\tau}$ as key optimization criteria over other parameters. The inverse relation is found for $FSRV$ and $RSRV$, which need to be minimized in solar cells. The extracted device parameters provide the experimentalist with a snapshot of the physical priorities at the optimal performance conditions; with this information, the experimentalist can focus in controlling or improving a specific device parameter, or performing secondary characterization to unveil further root-causes of underperformance.

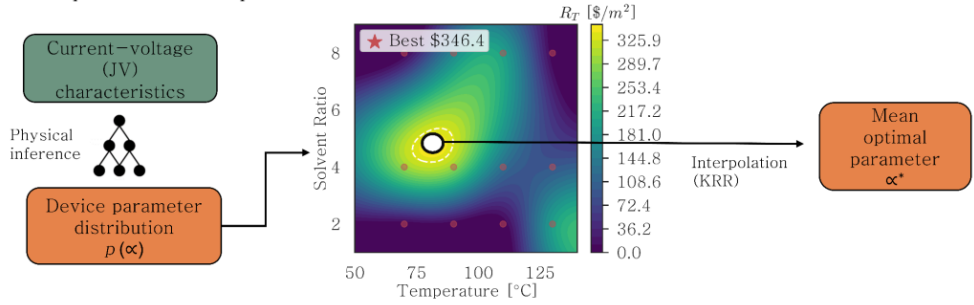
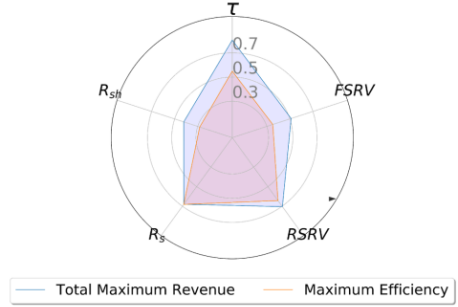

**Figure 5: Inference of mean optimal device parameters.** A. The current-voltage characteristics are used to extract a distribution of device parameters p($\alpha$) by scalable physical inference. Then, we compute the mean device parameter for each process condition, and use kernel ridge regression (KRR) to interpolate the value of the device parameters at the optimum. B. Normalized combinations of mean optimal device parameters at the optimum of two different figures of merit.

**Discussion and Closing Remarks**

In this perspective, we propose a data-driven optimization framework to bridge the gap between photovoltaics and industrial manufacturing. Our framework addresses two major causes of lengthy technology transfer: 1) Mismatch between figures of merit of R&D and industry, 2) Non-scalability of physical hypothesis testing and underperformance root-cause analysis. To accomplish this, our framework combines a technoeconomic figure of merit and scalable physical inference.

For our technoeconomic optimization objective, we propose total revenue $R_T$ as a consistent figure of merit for both R&D and industrial manufacturing. $R_T$ favors high efficiency of a single or few samples in early-stage R&D, while factoring yield and variability considerations into late stage research. This allows to interpret the transtion of R&D to industrial manufacturing as a continuum of economic value: high efficiency outliers have most of the value in early-stage R&D, but as we transition into industrial manufacturing, above certain efficiencies, yield and variability start having significant value. Our case study demonstrates the effectiveness of $R_T$. In the case of mature MAPbI$_3$ technology, we argue that using a similar manufacturing-relevant figure of merit can be essential for industrial scaling.

In our case study, we propose minimum sustainable price and market-competitive efficiency as parameters of $R_T$. In mature markets, a efficiency-revenue approach based on iso-LCOE can be more appropriate[31]. Furthermore, more complex technoeconomic constraints, solar cell degradation or manufacturability[47,48], can be included explicitly in the optimization problem. A related consideration is variability, and the amount of data available for the optimization. In general, the optimization objective $R_T$, as it is computed as a mean statistic of the revenue $R$, is not necessarily robust to outliers. In this case, an alternative approach relies on using robust statistics such L-estimators or M-estimators[49]. Similarly, surrogate-based optimization aims to capture the functional relation between the objective and the optimization variables. For this purpose, although pure black-box surrogate models can be used such as Gaussian process regression[40] or Bayesian neural networks[41], physics-informed surrogate models[23,39] can be useful in cases where enough information about the process physics, in the form of simulation or theoretical priors, is available. Finally, heteroskedasticity can play a significant role during closed-loop optimization, thus we recommend techniques such as heteroskedastic or deep GPR.

For scalable device-parameter inference, we demonstrated the inference of device parameters for numerous cells. This physical-granularity allows to map process parameters to optimal conditions, as shown in the perovskites case study, or perform a targeted optimization loop for device parameters (Eq. 6)[23]. The use of surrogate machine learning models, trained on physical simulations, along with rapid electrical characterization make inference scalable.

Since the inference procedure requires a device-physics model, model calibration with experimental data is crucial. Parameter inference often requires a carefully chosen set of measurements to decouple the contribution of competing physical parameters, and avoid multi-modal or ill-defined problems. For instance, current-voltage measurements can be performed at different illumination intensities and temperatures to decouple related solar cell parameters, such as bulk lifetime and minority-carrier mobility. Moreover, as the variability between efficiencies becomes smaller or small batches are used, the signal-to-noise ratio diminishes, and the inference loses statistical power. In those cases, having additional measurements, such as external quantum efficiency (EQE), can be effective to further decouple physical parameters and perform hypothesis testing. Thus, it is recommended that the experimentalist identifies the dominant material parameters for a certain problem, and defines the series of measurements required to decouple interacting parameters.

In the future, we envision our approach as part of accelerated photovoltaics development. Our framework can be seamlessly adapted to the fast cycles of high-throughput experimentation, allowing end-to-end solar cell optimization. The framework can also be extended to other similar aggregated material systems, such as batteries or transistors. As accelerated experimentation matures, we expect the gap in performance between photovoltaics R&D and industrial manufacturing to narrow, thanks to the introduction of multi-objective figures of merit and physical inference.

**Acknowledgements:**

We thank Jose Dario Perea (U. Toronto), Liu Zhe (MIT), Juan-Pablo Correa-Baena (Georgia Tech), Joel Jean (Swift Solar), Markus Gloeckler (First Solar), and Bill Huber (First Solar) for helpful discussions. This work was supported by a TOTAL SA research grant funded through MITei (supporting the experimental XRD), the National Research Foundation (NRF), Singapore through the Singapore Massachusetts Institute of Technology (MIT) Alliance for Research and Technology's Low Energy Electronic Systems research program, and by the U.S. Department of Energy under the Photovoltaic Research and Development program under Award DE-EE0007535. C.J.B. gratefully acknowledges the financial support through the "Aufbruch Bayern" initiative of the state of Bavaria (EnCN and SFF), the Bavarian Initiative "Solar Technologies go Hybrid" (SolTech), and DFG SFB953 (project no. 182849149) and DFG INST 90/917-1 FUGG.

**Author Contributions**

FO, ZR, JF, TB conceived the research and wrote the paper with inputs from all co-authors. MY, KZ, LN, TH performed perovskites synthesis and characterization. HX performed perovskite simulations. FO, ZR, SIPT performed machine learning analysis and calculations. SS, IMP, CB, BM, EB contribued with intellectual discussion.

**Declaration of Interests**
The authors declare no competing interests.

**Data availability**
The data and code used for this work is available in: https://github.com/PV-Lab/Data-Driven-PV